\documentclass[twoside,twocolumn,9pt]{article}
\usepackage{extsizes}
\usepackage[super,sort&compress,comma]{natbib} 
\usepackage[version=3]{mhchem}
\usepackage[left=1.5cm, right=1.5cm, top=1.785cm, bottom=2.0cm]{geometry}
\usepackage{balance}
\usepackage{mathptmx}
\usepackage{amsmath}
\usepackage{bm}
\usepackage{amssymb}
\usepackage{sectsty}
\usepackage{graphicx} 
\usepackage{lastpage}
\usepackage[format=plain,justification=justified,singlelinecheck=false,font={stretch=1.125,small,sf},labelfont=bf,labelsep=space]{caption}
\usepackage{float}
\usepackage{fancyhdr}
\usepackage{fnpos}
\usepackage[english]{babel}
\addto{\captionsenglish}{%
  
}
\usepackage{array}
\usepackage{droidsans}
\usepackage{charter}
\usepackage[T1]{fontenc}
\usepackage[usenames,dvipsnames]{xcolor}
\usepackage{setspace}
\usepackage[compact]{titlesec}
\usepackage{hyperref}
\usepackage{float}
\usepackage{upgreek}

\usepackage{epstopdf}

\definecolor{cream}{RGB}{222,217,201}

\begin{document}

\pagestyle{fancy}
\thispagestyle{plain}
\fancypagestyle{plain}{
\renewcommand{\headrulewidth}{0pt}
}

\makeFNbottom
\makeatletter
\renewcommand\LARGE{\@setfontsize\LARGE{15pt}{17}}
\renewcommand\Large{\@setfontsize\Large{12pt}{14}}
\renewcommand\large{\@setfontsize\large{10pt}{12}}
\renewcommand\footnotesize{\@setfontsize\footnotesize{7pt}{10}}
\makeatother

\renewcommand{\thefootnote}{\fnsymbol{footnote}}
\renewcommand\footnoterule{\vspace*{1pt}%
\color{cream}\hrule width 3.5in height 0.4pt \color{black}\vspace*{5pt}} 
\setcounter{secnumdepth}{5}

\makeatletter 
\renewcommand\@biblabel[1]{#1}            
\renewcommand\@makefntext[1]%
{\noindent\makebox[0pt][r]{\@thefnmark\,}#1}
\makeatother 
\renewcommand{\figurename}{\small{Fig.}~}
\sectionfont{\sffamily\Large}
\subsectionfont{\normalsize}
\subsubsectionfont{\bf}
\setstretch{1.125} 
\setlength{\skip\footins}{0.8cm}
\setlength{\footnotesep}{0.25cm}
\setlength{\jot}{10pt}
\titlespacing*{\section}{0pt}{4pt}{4pt}
\titlespacing*{\subsection}{0pt}{15pt}{1pt}

\fancyfoot{}
\fancyfoot[RO]{\footnotesize{\sffamily{1--\pageref{LastPage} ~\textbar  \hspace{2pt}\thepage}}}
\fancyfoot[LE]{\footnotesize{\sffamily{\thepage~\textbar\hspace{3.45cm} 1--\pageref{LastPage}}}}
\fancyhead{}
\renewcommand{\headrulewidth}{0pt} 
\renewcommand{\footrulewidth}{0pt}
\setlength{\arrayrulewidth}{1pt}
\setlength{\columnsep}{6.5mm}
\setlength\bibsep{1pt}
\newcommand{\forus}[1]{\textcolor[rgb]{0.80,0.20,0.00}{#1}}
\newcommand{\redtext}[1]{\textcolor[rgb]{0.80,0.20,0.00}{#1}}
\newcommand{\response}[1]{\textcolor[rgb]{0.00,0.00,1.00}{#1}}

\makeatletter 
\newlength{\figrulesep} 
\setlength{\figrulesep}{0.5\textfloatsep} 

\newcommand{\topfigrule}{\vspace*{-1pt}%
\noindent{\color{cream}\rule[-\figrulesep]{\columnwidth}{1.5pt}} }

\newcommand{\botfigrule}{\vspace*{-2pt}%
\noindent{\color{cream}\rule[\figrulesep]{\columnwidth}{1.5pt}} }

\newcommand{\dblfigrule}{\vspace*{-1pt}%
\noindent{\color{cream}\rule[-\figrulesep]{\textwidth}{1.5pt}} }

\makeatother

\twocolumn[
  \begin{@twocolumnfalse}
\vspace{1em}
\sffamily
\begin{tabular}{m{2.5cm} p{15.5cm} }

& \noindent\LARGE{\textbf{A theory of ordering of elongated and curved proteins on membranes driven by density and curvature}} \\
\vspace{0.3cm} & \vspace{0.3cm} \\

 & \noindent\large{Caterina Tozzi,\textit{$^{a\dag}$} Nikhil Walani,\textit{$^{a\dag}$}, Anabel-Lise Le Roux{$^{b}$}, Pere Roca-Cusachs{$^{bc}$},  and Marino Arroyo\textit{$^{\star abd}$}} \\

\\

& \noindent\normalsize{
Cell membranes interact with a myriad of curvature-active proteins that control membrane morphology and are responsible for mechanosensation and mechanotransduction. Some of these proteins, such as those containing BAR domains, are curved and elongated, and hence may adopt different states of orientational order, from isotropic to maximize entropy to nematic as a result of crowding or to adapt to the curvature of the underlying membrane. Here, extending the work of [Nascimento et. al, \textit{Phys. Rev. E}, 2017, 96, 022704], we develop a mean-field density functional theory to predict the orientational order and evaluate the free-energy of ensembles of elongated and curved objects on curved membranes. This theory depends on the microscopic properties of the particles and  explains how a density-dependent isotropic-to-nematic transition is modified by anisotropic curvature. We also examine the coexistence of isotropic and nematic phases. This theory lays the ground to understand the interplay between membrane reshaping by BAR proteins and molecular order, examined in [Le Roux et. al, \textit{Submitted}, 2020].} \\

\end{tabular}

 \end{@twocolumnfalse} \vspace{0.6cm}

  ]

\renewcommand*\rmdefault{bch}\normalfont\upshape
\rmfamily
\section*{}
\vspace{-1cm}


\footnotetext{\textit{$^{a}$~Universitat Polit\`ecnica de Catalunya-BarcelonaTech, 08034 Barcelona, Spain. Email: marino.arroyo@upc.edu}}
\footnotetext{\textit{$^{b}$~Institute for Bioengineering of Catalonia (IBEC), The Barcelona Institute for Science and Technology (BIST), 08028 Barcelona, Spain. }}

\footnotetext{\textit{$^{c}$~Universitat de Barcelona, 08036 Barcelona, Spain.}}

\footnotetext{\textit{$^{d}$~Centre Internacional de M\`etodes Num\`erics en Enginyeria (CIMNE), 08034 Barcelona, Spain. }}
\footnotetext{\textit{$^{\dag}$~These authors contributed equally to this work.}}


Many cellular processes rely on the ability of cell membranes to change their shape, including area and tension regulation \cite{Sinha2011}, or the transport of cargo within the cells in membrane bound vesicles \cite{Bonifacino2004}. Membrane can change shape in response to cytoskeletal dynamics \cite{Tinevez2009}, changes in pH of the surrounding medium \cite{Khalifat2014}, or the recruitment of proteins that are either curved \cite{Zimmerberg2006} or bulky and disordered \cite{Stachowiak2012}. In this study we focus on elongated curved proteins such as those containing the BAR domain (amphiphysin, endophilin, F-CHO) and others like dynamin, EHD2, etc \cite{LeRoux19}. When these proteins lack positional order but tend to point in a given direction, at least locally, they are in a so-called nematic state. These proteins are ``banana shaped'' and can impinge anisotropic curvatures on the membranes upon binding through a scaffolding effect \cite{Kabaso2011}, which allows them to tubulate liposomes\cite{Peter2004}, stabilize tubular necks in Caveolae \cite{Hoernke2017} or bind to necks of budding vesicles and drive endocytic transport \cite{Liu2011}. The generation of anisotropic curvature has been associated with a nematic ordering of the elongated proteins along the high-curvature direction at very high coverage \cite{Frost2008,Mim2012}. Besides anisotropic curvature, elongated proteins can also create isotropically curved (spherical) domains, as F-BARs in the initial stages of assembly of clarthin coats \cite{Henne2010} or during fast endocytosis by endophilin \cite{Boucrot2015}. This suggests a multi-functionality of curved and elongated proteins and a correlation between curvature anisotropy, density, and nematic order. Controlled in-vitro experiments  capturing the dynamics of this interplay have been elusive. Giant Unilamellar Vesicles (GUVs) exposed to curved and elongated proteins exhibit no change in membrane shape below a tension-dependent protein coverage threshold, above which very thin protein-rich tubules are violently shed by the vesicle  \cite{Shi2015}, and in GUVs-tether systems, the high membrane tension strongly reduces the ability of the membrane to change shape \cite{Zhu2012,Sorre2012}. 

Theoretical continuum models coupling membrane's elasticity with orthotropic proteins \cite{Kabaso2012, Walani2014} have been restricted to a prescribed configuration (density and orientation) of proteins that are always in the nematic phase. Coarse grained MD and Monte-Carlo simulations on the other hand have been successfully used to understand molecular aspects of proteins in shaping of membrane such as scaffolding as compared to helical insertions \cite{Arkhipov2009,Yu2013,Mim2012}, arc length \cite{Bonazzi2019}, lateral interactions or chirality \cite{Noguchi2016, Noguchi2019} and the generation of topological defects in  closed vesicles transforming into tubular liposomes \cite{Ramakrishnan2014}. However, these simulations are limited to short time-scales and small length-scales. While continuum models have shown promise in describing the dynamics of membrane protein interactions at time and length scales relevant to many biophysical processes \cite{Tozzi2019,ArroyoOnsager,Sahu2017,Agrawal2011,Mahapatra2020}, they are phenomenological and disconnected to the relevant microscopic details. Thus, there is a need for the development of effective field theories  capturing the microscopic details of protein interactions on lipid membranes. 

To this end, we develop a mean field density functional theory for the free energy of the proteins  accounting for protein area coverage, orientational order and membrane curvature that can provide a basis for upscaling. In the present work, membrane curvature is taken as given. In a companion paper, we couple the model presented here with one for a deformable membrane to study the mechano-chemistry of membrane reshaping by BAR proteins \cite{LeRoux20}. In Section 1, the entropy of elongated proteins on the membrane is obtained by adapting to 2D a recent theory by Nascimento et. al \cite{Nascimento_2017} for hard ellipsoidal particles, which corrects Onsager's classical theory of isotropic-to-nematic transitions for non-spherical particles to provide quantitative prediction at high densities and moderate particle aspect ratio. The crucial difference with Onsager's work \cite{Onsager1949} is the enforcement of a compact support of the  orientational probability distribution  beyond a certain spatial density. In Section 2, we generalize the theory to account for the elastic curvature energy of the proteins depending on their orientation and on the second fundamental form of the underlying surface. We also examine the effect of curvature on the isotropic-to-nematic transition and on the orientational probability distribution. Finally, in Section 3, we study the coexistence of isotropic and nematic phases. 

\section{Configurational free-energy of proteins as elliptical particles}

\subsection{Mean field approximation}

\begin{figure*}
 \centering
\includegraphics[width=7in]{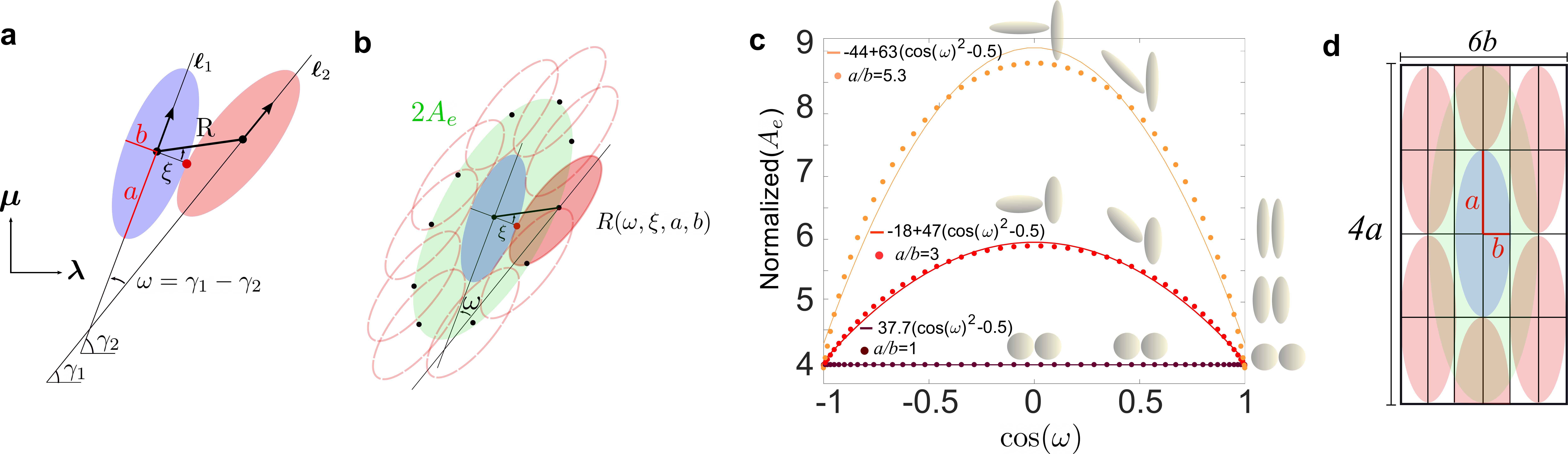}
\caption{\label{Orient_Ellipse} (a) Contact configuration of two ellipses in the plane. The modulus of the vector joining the centers, $R = \vert\bm{R}\vert$, is the so-called distance of closest approach for two ellipses whose major axis forms an angle $\omega = \gamma_1-\gamma_2$, where $\gamma_\alpha$ is the orientation of each of the ellipses with respect to a fixed direction. This distance obviously depends on $\omega$, on the length of the major and minor axes, $2a$ and $2b$, but also on the location of the contact point, parametrized by the angle $\xi$. (b) Illustration of the calculation of the excluded area (shaded in green) for a given $\omega$. (c) Excluded area per particle for different ellipses of fixed area as a function of $\cos \omega$. The excluded area is normalized by the area of an ellipse $\pi a b$ and all the ellipses with different aspect ratios $a/b$ have the same area. Elongated ellipses need to align to reduce the excluded area per ellipse. Dots represent the excluded area computed numerically with high accuracy while the solid line corresponds to a least-squares fit with the second-order polynomial approximation in Eq.~\eqref{Approx_Area}. (d) Schematic view for the choice of $\lambda$ in Eq.~\eqref{W_excluded} estimated through the relation $a_{\rm eff}= \lambda A_e$ \cite{Nascimento_2017}, where $a_{\rm eff}$ is the average area effectively occupied by one particle in the rectangle and given by $a_{eff}=A_{rect}/N_{\rm particles}=4ab$ and $A_e$ is the excluded area for a pair of particles with parallel long axis, $A_e = 4 \pi ab $. }
\end{figure*}

Before accounting for the bending energy of proteins, we adapt and extend the theory presented by \cite{Nascimento_2017} to two-dimensions to apply it to proteins on a membrane. We consider proteins to be 2D elliptical particles such that the length of their major and minor axes are given by $2a$ and $2b$ as shown in Fig.~\ref{Orient_Ellipse}.  The state of protein $i$, $\bm{q}_{i}$, is given by its position on a surface $\Gamma$, $\bm{r}_i$ and orientation over the unit circle $\mathbb{S}$, given by the angle $\gamma_i$ of the long axis of the ellipse relative to a fixed direction on the surface. We assume that proteins are rigid, non-overlapping but otherwise non-interacting. Thus, for two particles with states $\bm{q}_1$ and $\bm{q}_2$, their interaction potential $U$ is purely repulsive and can be defined as 
\begin{equation}\label{Int_Potential}
U(\bm{q}_1,\bm{q}_2) = \begin{cases}
\infty &\text{if particles overlap,}\\
0 &\text{otherwise}.
\end{cases}
\end{equation}
The configurational free energy for $N$ identical proteins up to additive constant is given by 
\begin{equation}
\mathcal{F}_{e} = - \frac{1}{\beta} \ln {\frac{1}{N!} \int_{\Omega^N} e^{-\beta \Sigma_{1 \leq i < j \leq N} U_{i,j}} \, d \bm{q}_1 \dots d \bm{q}_N},
\end{equation}
where $\beta = 1/k_B T$ and $\Omega$ is the domain in phase space for each of the proteins, accounting for the translational and orientational degrees of freedom. Since all the proteins are equivalent, invoking a mean field approximation and a passage to the continuum limit (see \cite{Nascimento_2017} for further details), the free energy can be written as 
\begin{equation}\label{Free_Energy_1}
\begin{aligned}
\mathcal{F}_{e} &= \frac{1}{\beta} \int_{\Omega} \rho(\bm{q}) \ln (\rho(\bm{q})) \, d\bm{q} - \frac{1}{\beta} \int_{\Omega} \rho(\bm{q}) \ln [1 - W(\bm{q})] \, d\bm{q}
\end{aligned}
\end{equation}
where $\rho(\bm{q})$ is the number density of particles with state $\bm{q}$ and $W(\bm{q})$ is the average fraction of excluded area for a given particle, i.e.~the fraction of phase space inaccessible to a particle due to presence of other particles. This quantity is postulated to take the form\cite{Nascimento_2017}
\begin{equation}\label{W_excluded}
W(\bm{q}_1) = \lambda \int_{\Omega} \rho(\bm{q}_2) \left[ 1 - e^{-\beta U(\bm{q}_1,\bm{q}_2)} \right] \, d\bm{q}_2,
\end{equation}
where $\lambda$ is an adjustable parameter discussed in Fig.~\ref{Orient_Ellipse} accounting for the high-density packing. We note that in the dilute limit, the first term in Eq.~(\ref{Free_Energy_1}) is dominant, and thus in this limit $\lambda$ does not play an important role. We further express the number  density of proteins in terms of positional and orientational contributions
\begin{equation} \label{separation}
\rho(\bm{q}) = \phi(\bm{r}) f(\bm{r},\gamma),
\end{equation}
so that $\int_{\mathbb{S}} f(\bm{r},\gamma) d\gamma = 1$ and $\int_{\Gamma} \phi(\bm{r}) d\bm{r} = N$. Substituting the above relation into Eq.~\eqref{W_excluded}, we obtain 
\begin{align}
W(\bm{q}_1) = \lambda \int_{\mathbb{S}} \int_{\Gamma} \phi(\bm{r}_2)f(\bm{r}_2,\gamma_2) \left[ 1 - e^{-\beta U(\bm{r}_1, \gamma_1,\bm{r}_2,\gamma_2)} \right]  \, d\bm{r}_2 \, d\gamma_2,
\end{align}
where we note that $d\bm{r}_2$ should be interpreted as the element of area on the surface $\Gamma$.
Note that the term between square brackets is zero unless $\bm{r}_1$ and $\bm{r}_2$ are within a small distance commensurate to the particle size. Thus, assuming that $\phi$ varies slowly in this length-scale, it is reasonable to approximate $\phi(\bm{r}_2)f(\bm{r}_2,\gamma) \approx \phi(\bm{r}_1)f(\bm{r}_1,\gamma)$ in the equation above, finding 
\begin{align} \label{W_app}
W(\bm{q}_1) =  \lambda \phi(\bm{r}_1) \int_{\mathbb{S}} f(\bm{r}_1,\gamma_2) A_{\text{e}}(\gamma_1,\gamma_2) \, d \gamma_2,
\end{align}
where $A_{\text{e}}(\gamma_1,\gamma_2)$ is given by 
\begin{align}
A_{\text{e}}(\gamma_1,\gamma_2) = \int_{\Gamma}\left[ 1 - e^{-\beta U(\bm{r}_1, \gamma_1,\bm{r}_2,\gamma_2)} \right]  \, d\bm{r}_2.
\end{align}
The integrand in this expression is 0 unless the two particles overlap, in which case it is 1. It thus contains purely geometric information and can be interpreted as the excluded area per particle for two particles oriented along the angles $\gamma_1$ and $\gamma_2$. Note that, by translational invariance, it is independent of $\bm{r}_1$, and by rotational invariance is should depend on $\gamma_1$ and $\gamma_2$ through their difference.

Introducing Eqs.~(\ref{separation},\ref{W_app}) into Eq.~\eqref{Free_Energy_1}, we can write the configurational free energy of the system as 
\begin{equation}\label{Free_Energy_2}
\begin{aligned}
\mathcal{F}_e =  & \frac{1}{\beta}\int_{\Gamma} \phi(\bm{r}) \ln \phi(\bm{r}) \, d\bm{r} 
\\ 
 + &\frac{1}{\beta} \int_{\Gamma} \phi(\bm{r}) \left\{ \int_{\mathbb{S}} f(\bm{r},\gamma) \left[ \ln f(\bm{r},\gamma) - \ln g(\bm{r},\gamma) \right]\, d\gamma \right\} \, d\bm{r}, 
\end{aligned}
\end{equation}
where we have defined
\begin{equation}
\begin{aligned}
g(\bm{r},\gamma) & = 1 - \lambda \phi(\bm{r}) \int_{\mathbb{S}} f(\bm{r},\gamma_2) A_{\text{e}}(\gamma,\gamma_2) \, d\gamma_2.
\end{aligned}
\end{equation}
We note that for circular particles, this free energy reduces to that of a Van der Waals gas. We also note that, even though $A_{\text{e}}$ is scale dependent (it has units of area), $\phi$ is also scale dependent in such a way that $g$ is dimensionless and scale-independent.

\subsection{Excluded area for two ellipses}

To evaluate the free-energy in Eq.~(\ref{Free_Energy_2}), we need to evaluate $A_{\text{e}}(\gamma_1,\gamma_2)$. For this, we note that the excluded area between  two ellipses can be computed in terms of the distance of closest approach $R$, see Fig.~\ref{Orient_Ellipse}(a,b), as\cite{PhysRevE.75.061709}  
\begin{align}
A_{\text{e}}(\gamma_1, \gamma_2) = \frac{1}{2} \int_{\mathbb{S}} R^2 (\omega,\xi, a, b) \, d\xi.
\end{align}
The calculation of $R(\omega,\xi, a, b)$ is algebraically complex\cite{Vieillard_72}. It can be shown that $R$ solves the equation 
\begin{align}
0 = 4(f_1^2 - 3 f_2)(f_2^2 - 3 f_1) - (9- f_1 f_2)^2,
\end{align}
where 
\begin{align}
f_\alpha = 1 + G - \left(\frac{R}{a} \sin\theta_\alpha\right)^2 - \left(\frac{R}{b} \cos\theta_\alpha\right)^2,~~~\mbox{for}~~~\alpha = 1,2,
\end{align}
with $\theta_1 = \xi$, $\theta_2 = \xi-\omega$, and 
\begin{align}
G = 2 + \left( \frac{a}{b} - \frac{b}{a} \right)^2 \sin^2{\omega}.
\end{align}
The above equations allow us to compute $A_{\text{e}}(\gamma_1, \gamma_2)$ numerically, see Fig.~\ref{Orient_Ellipse}(c), which shows how the dependence of the excluded area on particle alignment depends on the aspect ratio. 

The average fraction of excluded area for a given particle $W(\bm{q})$ is only relevant at high packing but we are estimating it using the excluded area between two particles, $A_{\text{e}}$. To reconcile these two quantities through the parameter $\lambda$, we follow  \cite{Nascimento_2017}. In a dense packing limit, such as in Fig.~\ref{Orient_Ellipse}(d), $W(\bm{q})$ should approach 1, which we can express as the number density of particles $\phi = N_{\rm particles}/A_{\rm tot}$ times an effective area per particle $a_{\rm eff}$ in such a dense arrangement. We thus obtain $a_{\rm eff} = 4ab$. Examining Eq.~(\ref{W_app}), in this high-packing limit $1\approx \lambda (N_{\rm particles}/A_{\rm tot}) A_{\rm e}$, where $A_{\rm e}$ is the excluded area between two ellipsoidal particles with parallel long axis, i.e.  $A_{\rm e} = 4\pi ab$. We thus conclude that $\lambda = 1/\pi \approx 1/3$.

Since $A_{\text{e}}(\gamma_1, \gamma_2)$ depends on the orientations of the particles through $\omega = \gamma_1 - \gamma_2$, it can be approximated by an expansion of Legendre polynomials depending on $\cos\omega$ as 
\begin{align} \label{Approx_Area}
A_{\text{e}}(\gamma_1, \gamma_2) = B_0 + B_2 P_2(\cos {\omega}) + \cdots
\end{align}
where $B_0$ and $B_2$ are constants depending on $a$ and $b$, and $P_2(x) = x^2  - 1/2$. Note that by symmetry arguments, only even polynomials appear in the expansion. The expansion can be extended to higher order but the second-order approximation already provides a good approximation, see Fig.~\ref{Orient_Ellipse}(c). Interestingly, the second order expansion allows us to express $A_{\text{e}}(\gamma_1, \gamma_2)$ in terms of the symmetric and traceless tensor
\begin{equation} \label{local_Q}
\bm{\sigma} (\gamma) = \frac{1}{2} \left[2 {\bm{\ell}} (\gamma) \otimes {\bm{\ell}} (\gamma) - \bm{I} \right],
\end{equation}
where $\bm{I}$ is the surface identity and $\bm{\ell}(\gamma)$ is the local orientation of proteins. The latter can be expressed in an arbitrary orthonormal frame of the tangent plane to the surface $\{\bm{\lambda},\bm{\mu}\}$ as $\bm{\ell} = \cos\gamma \,\bm{\lambda} +  \sin\gamma \,\bm{\mu}$, see Fig.~\ref{Orient_Ellipse}. This tensor describes the local (or microscopic) second moment of the orientation of proteins, and as shown later, it leads to a theory where orientational order appears through the classical nematic tensor $\bm{Q}$.

By noting that $\bm{\ell}(\gamma_1)\cdot \bm{\ell}(\gamma_2) = \cos\omega$, a direct calculation shows from Eq.~(\ref{Approx_Area}) that
\begin{equation} \label{Aux_aspect}
\lambda A_{\text{e}}(\gamma_1, \gamma_2) = c - d \bm{\sigma} (\gamma_1) :  \bm{\sigma} (\gamma_2),
\end{equation}
where, $c= \lambda B_0$ and $d= - \lambda B_2$ depend on $a$ and $b$ and can be computed by fitting a second order polynomial in $\cos\gamma$ to $A_{\text{e}}(\gamma_1, \gamma_2)$.

\subsection{Optimizing the orientational distribution}

Given a density field $\phi$, we can find the optimal angular distribution at each point in space $\bm{r}$ by minimizing the free energy in Eq.~\eqref{Free_Energy_2} with respect to $f$ subject to the normalization constraint.  To minimize the free energy and account for this constraint, we introduce the Lagrangian functional 
\begin{align}
\mathcal{L}[f,\mu] = \int_{\mathbb{S}} f \left(\ln f-\ln g \right)d\gamma +  \mu \left(\int_{\mathbb{S}} f \, d\gamma - 1 \right),
\end{align}
where $\mu$ is a Lagrange multiplier  field. 
Since we perform this minimization point-wise, we drop for notational simplicity the dependence on $\bm{r}$ of $f$, $g$, $\mu$, $\phi$ and all quantities depending on these fields.

Recalling Eq.~(\ref{Aux_aspect}), we have
\begin{equation}\label{g_def}
\begin{aligned}
g(\gamma) = 1-\phi (c   - d  \bm{\sigma} (\gamma): \bm{Q}),
\end{aligned}
\end{equation}
where we have introduced the nematic tensor describing the average particle orientation
\begin{equation} \label{Def_Q}
\bm{Q} = \int_{\mathbb{S}} f(\gamma) \bm{\sigma} (\gamma) \, d\gamma = \left< \bm{\sigma} (\gamma) \right>.
\end{equation}
Note that $\bm{Q}$ inherits from $\bm{\sigma}$ the properties of being symmetric and traceless. We further introduce the auxiliary tensor 
\begin{equation} \label{Aux_def}
\bm{\psi} = d \phi \int_{\mathbb{S}} \frac{f(\gamma)}{g(\gamma)} \bm{\sigma} (\gamma) \, d\gamma,
\end{equation}
which is also symmetric and traceless. The stationarity condition can then be written as 
\begin{align}
0 = \delta_f \mathcal{L} & = \int_{\mathbb{S}} \left(\ln f - \ln g -  \bm{\psi}:\bm{\sigma} + \mu \right) \delta f d\gamma,
\end{align}
for all admissible variation $\delta f$, and thus the term between parentheses must vanish. It is clear that when $g\rightarrow 0$, then necessarily $f\rightarrow 0$. We can thus define the support of $f(\gamma)$ as $\mathbb{S}^+ = \{\gamma \in (-\pi, \pi)~\mbox{such that}~g(\gamma)>0\}$. Determining $\mu$ though the normalization of $f$, we find an expression for the angular probability density function
\begin{equation} \label{PDF_1}
f(\gamma) = \begin{cases}
 \dfrac{g(\gamma) e^{\bm{\sigma}(\gamma) :  \bm{\psi}}}{ \int_{\mathbb{S}^+} g(\gamma') e^{\bm{\sigma}(\gamma') :  \bm{\psi}} \, d\gamma'} & \text{if } \gamma\in \mathbb{S}^+ \\
0  & \text{otherwise.}
\end{cases}
\end{equation}
We note that this expression is far from being explicit, since $g$ depends on $\bm{Q}$, which in turn depends on $f$, and $\bm{\psi}$ also depends on $f$. However, as developed below, it allows us to evaluate the free energy. We also note that the probability density function $f$ vanishes in a region of the orientational space as the areal number density of proteins $\phi$ increases, and thus $g$ in Eq.~(\ref{g_def}) becomes negative. As discussed in \cite{Nascimento_2017,Taylor2018}, this is critical to quantitatively predict density based ordering for moderately elongated particles. Finally, due to the symmetry of particles with respect to rotations by $\pi$, it follows that $f(\gamma) = f(\gamma + \pi)$.

Since $\bm{Q}$ is symmetric and traceless, it has two real eigenvalues of opposite sign and it is diagonal in an orthonormal eigenframe. We let $\{\bm{\lambda},\bm{\mu}\}$ be this eigenframe. Thus, the nematic tensor can be expressed as 
\begin{equation}\label{Q_eig}
\bm{Q} = \frac{S}{2} \left( \bm{\lambda} \otimes \bm{\lambda} - \bm{\mu} \otimes \bm{\mu} \right),
\end{equation}
where we call $S$ the order parameter, which contracting the above relation  and Eq.~(\ref{Def_Q}) with $\bm{\lambda} \otimes \bm{\lambda}$ can be expressed as 
\begin{equation} \label{Order_Def}
S = \left< 2 \left( \bm{\ell} \cdot \bm{\lambda} \right)^2 - 1 \right> = 2 \left< P_2(\cos \gamma) \right>,
\end{equation}
where $\gamma$ is the angle between the nematic direction $\bm{\lambda}$ and the direction of a microscopic particle, $\bm{\ell}$. Combining Eqs.~(\ref{local_Q}) and (\ref{Q_eig}), we also find that $\bm{\sigma}: \bm{Q}  = S P_2(\cos \gamma)$, and thus 
\begin{align}\label{g_expr}
g(\gamma) = 1-\phi [c   - d  S P_2(\cos \gamma)].
\end{align}

A traceless symmetric tensor such as $\bm{\psi}$ can be expressed in the eigenframe of $\bm{Q}$ as 
\begin{align}
\bm{\psi} = \frac{\psi}{2} (\bm{\lambda}\otimes\bm{\lambda} - \bm{\mu}\otimes\bm{\mu}) + \frac{\bar{\psi}}{2} (\bm{\lambda}\otimes\bm{\mu} + \bm{\mu}\otimes\bm{\lambda}).
\end{align}
We show next that in fact, $\{\bm{\lambda},\bm{\mu}\}$ is also an eigenfame of $\bm{\psi}$, and thus $\bar{\psi}= 0$.

With the above representation of $\bm{\psi}$, we find that 
\begin{align}\label{sigma_psi}
\bm{\sigma}:\bm{\psi} = \psi P_2(\cos\gamma) + \bar{\psi} \sin\gamma \cos\gamma.
\end{align}
The condition that $\{\bm{\lambda},\bm{\mu}\}$ is an eigenframe of $\bm{Q}$ implies that $0 = \bm{\lambda}\cdot \bm{Q} \cdot \bm{\mu}$ and hence
\begin{align}\label{condx}
0 = \int_{\mathbb{S}^+} f(\gamma)\bm{\lambda}\cdot \bm{\sigma}(\gamma) \cdot \bm{\mu} ~d\gamma = \int_{\mathbb{S}^+} f(\gamma)\sin\gamma \cos\gamma ~d\gamma.
\end{align}
Noting that $\mathbb{S}^+$ is symmetric about $\gamma = 0$ since $g(\gamma)$ is an even function, Eq.~(\ref{g_expr}), and using the symmetry $f(\gamma) = f(\gamma + \pi)$, the above relation implies that
 \begin{align*}
0 & =  \int_{\mathbb{S}^+\cap (-\pi/2, \pi/2)} g(\gamma) e^{\bm{\sigma}(\gamma):\bm{\psi}}  \sin\gamma \cos\gamma ~d\gamma \\
& = \int_{\mathbb{S}^+\cap (-\pi/2, \pi/2)} g(\gamma)  \sin\gamma \cos\gamma \, e^{\psi P_2(\cos\gamma)} e^{ \bar{\psi} \sin\gamma \cos\gamma}~d\gamma \\
& = \int_{\mathbb{S}^+\cap (0, \pi/2)}  g(\gamma)  \sin\gamma \cos\gamma \, e^{\psi P_2(\cos\gamma)} \left(e^{ \bar{\psi} \sin\gamma \cos\gamma} - e^{ -\bar{\psi} \sin\gamma \cos\gamma} \right) d\gamma,
\end{align*}
where in the last step we have used the fact that $g(\gamma)  \sin\gamma \cos\gamma \, e^{\psi P_2(\cos\gamma)}$ is an odd function of $\gamma$.
Since in the integration domain $\mathbb{S}^+\cap (0, \pi/2)$ the function $g(\gamma)$, $\sin\gamma$ and $\cos\gamma$ are strictly positive, it follows that the integral above is strictly positive if $\bar{\psi}>0$ and strictly negative if $\bar{\psi}<0$, and we thus conclude that $\bar{\psi} = 0$, that $\bm{Q}$ and $\bm{\psi}$ have the same eigenframe, and that $\bm{\sigma}:\bm{\psi} = \psi P_2(\cos\gamma)$.

\begin{figure}[hbtp!]
 \centering
\includegraphics[width=2.4in]{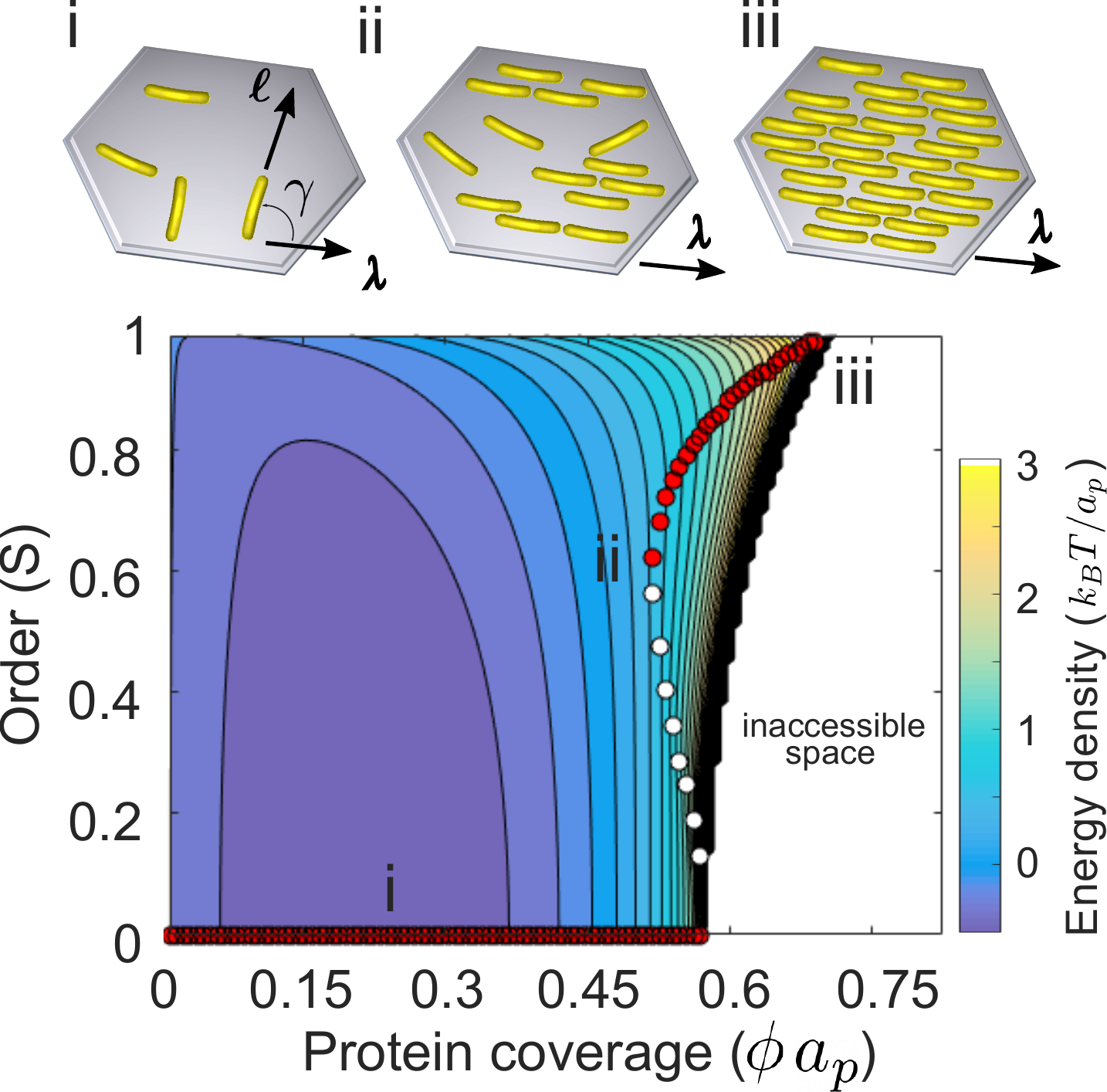}
\caption{\label{Energy_ellipses} Free-energy density landscape as a function of protein coverage, expressed as the area fraction $a_p \phi$ with $a_p$ the area of a protein, and of nematic order $S$. The white region is inaccessible due to crowding. There is a discontinuous isotropic-to-nematic transition for an area fraction of about 0.5. We consider ellipses with aspect ratio $a/b = 3$ on a flat membrane. Dots represent minima (red) and maxima (white) of the energy profile for fixed $\phi$. The diagrams on top illustrate  states i, ii and iii.}
\end{figure}

Combining this last expression with Eqs.~(\ref{PDF_1},\ref{Order_Def},\ref{g_expr}), we find that
\begin{align}\label{root}
S = 2\frac{\int_{\mathbb{S}^+} P_2(\cos\gamma) \left\{1-\phi\left[c - d S P_2(\cos\gamma)   \right] \right\} e^{\psi P_2(\cos\gamma)} d\gamma}{\int_{\mathbb{S}^+}  \left\{1-\phi\left[c - d S P_2(\cos\gamma)   \right] \right\} e^{\psi P_2(\cos\gamma)} d\gamma}.
\end{align}
Importantly, the above relation provides an implicit relation for the auxiliary variable $\psi(\phi,S)$ given the particle number density $\phi$ and the order parameter $S$. With these expressions, the configurational free-energy in Eq.~(\ref{Free_Energy_2}) can be rewritten as
\begin{equation}\label{Free_Energy_3}
\begin{aligned}
\mathcal{F}_e[\phi,S] =  & \frac{1}{\beta}\int_{\Gamma} \phi \bigg\{  \ln \phi + \frac{S \psi}{2}  \\ & - \ln\int_{\mathbb{S}^+} \left\{1-\phi\left[c - d S P_2(\cos\gamma)   \right] \right\} e^{\psi P_2(\cos\gamma)} d\gamma \bigg\}\, d\bm{r}, 
\end{aligned}
\end{equation}
in terms of the fields $\phi(\bm{r})$ and $S(\bm{r})$. We note that, given the lack of a preferred orientation, this effective free-energy depends on the nematic tensor $\bm{Q}$ only through $S$, and whenever $S\ne 0$, the nematic direction is arbitrary.

\begin{figure*}[h!]
 \centering
\includegraphics[width=6.5in]{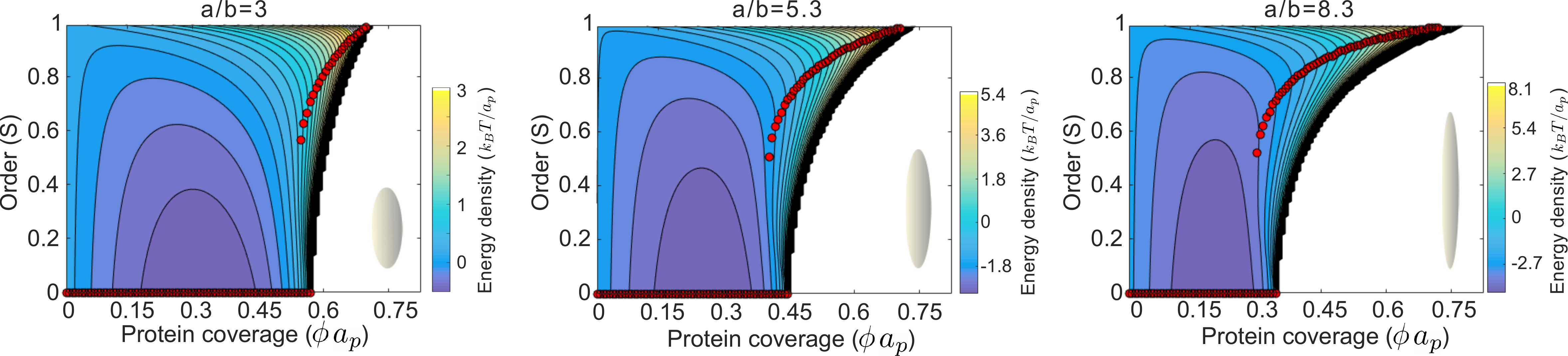}
\caption{\label{Energy_ellipses_aspect} Energy density landscape for ellipses with varying aspect ratio on a flat membrane. }
\end{figure*}

\subsection{Free-energy landscapes on planar surfaces}

Figure \ref{Energy_ellipses} shows the landscape of the energy density, the integrand in Eq.~(\ref{Free_Energy_3}), as a function of density and order parameter. In the figure, we express density or coverage as the area fraction $a_p\phi$, where $a_p$ is the area of a protein. The figure shows that, as area fraction becomes large, the energy grows rapidly irrespective of $S$ and blows up at finite density, defining a region of inaccessible states where $g(\gamma)$, see Eq.~(\ref{g_expr}), becomes  negative. Given the density $\phi$, we can minimize the energy profile with respect to $S$ to determine the degree of order in equilibrium as a function of $\phi$, defining the equilibrium path shown by red dots in the figure. For low area-fraction, proteins maximize their entropy by being randomly oriented and hence $S=0$ is the only solution branch. As density increases beyond a threshold, we observe the emergence of another stable branch characterized by high protein order. There is a range of densities where both the disordered and the ordered branches coexist and are separated by unstable equilibrium points marked in the figure with white dots. Density-based ordering for such elliptical molecules thus proceeds through a discontinuous phase transition. The procedure described here, which adapts that in \cite{Nascimento_2017} to 2D systems with elliptical particles, predicts how the energy landscape depends on the particle aspect ratio. As shown in Fig.~\ref{Energy_ellipses_aspect}, increasing it decreases the size of the region of accessible states, decreases the threshold density of the phase transition, and increases the maximum packing limit.

\begin{figure}[hbtp!]
 \centering
\includegraphics[width=3in]{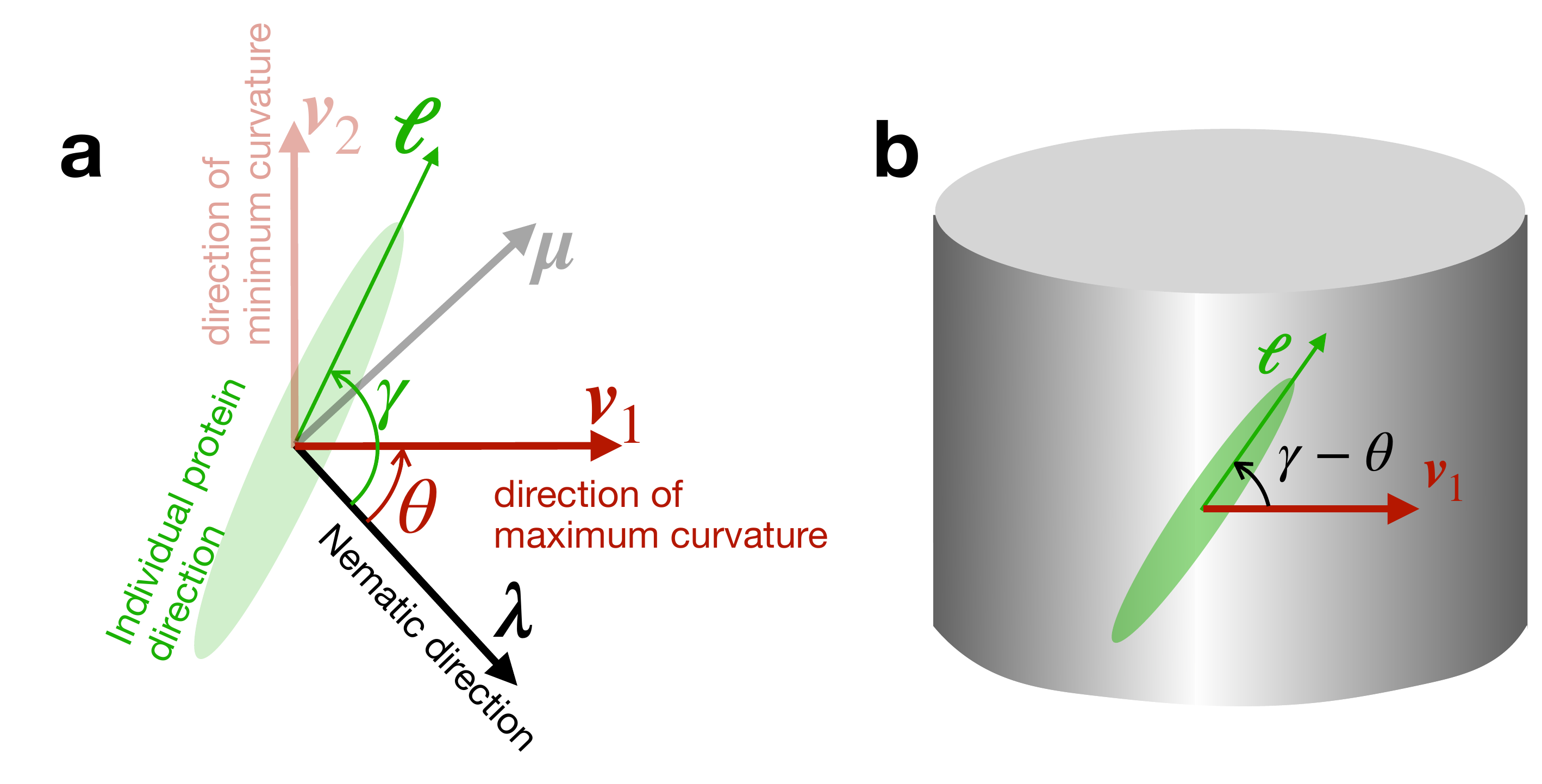}
\caption{\label{frames} (a) Illustration of the eigenframe $\{\bm{\lambda} , \bm{\mu} \}$ of the nematic tensor $\bm{Q}$, where $\bm{\lambda}$ is the nematic direction, and of the eigenframe $\{ \bm{v}_1,\bm{v}_2 \}$ of the second fundamental form of the surface $\bm{k}$, where these vectors determine directions of maximum and minimum curvature of the surface. We also illustrate a microscopic direction $\bm{\ell}$,  the angle $\gamma$ between the nematic direction $\bm{\lambda}$ and $\bm{\ell}$, and the angle $\theta$ between the two eigenframes. (b) An adsorbed protein along vector $\bm{\ell}$ samples the normal curvature of the surface in this direction.}
\end{figure}

\section{Free-energy of bendable proteins on curved membranes}

\subsection{Accounting for the bending energy}

Having described the phase transition of ellipses on flat surfaces, we now consider curved proteins adhered to a curved lipid membrane approximated as a surface. In doing so, we ignore the effect of curvature in the calculation of the excluded area but account for the bending energy of the proteins. Since binding of BAR proteins occurs due to electrostatic interaction with the lipids, they adhere along their charged faces  \cite{Zimmerberg2004,Kabaso2011}. We assume that adsorbed proteins sample the curvature of the underlying membrane along their long axis $\bm{\ell}$, Fig.~\ref{frames}(b), i.e.~the surface normal curvature along this direction, which can be computed as $k_{\bm{\ell}} = \bm{\ell}\cdot \bm{k} \cdot \bm{\ell}$ where $\bm{k}$ is the second fundamental form of the membrane surface characterizing its local curvature. 

Since both the nematic tensor $\bm{Q}$, see Eq.~(\ref{Def_Q}), and the second fundamental form of the surface are symmetric tensors, they possess respective tangential orthonormal eigenframes, $\{ \bm{\lambda},\bm{\mu} \}$ for $\bm{Q}$ and one given by the principal curvature directions $\{ \bm{v}_1,\bm{v}_2 \}$ for $\bm{k}$, Fig.~\ref{frames}(a). The principal curvatures of the surface are the corresponding eigenvalues $\bm{k}\cdot \bm{v}_i = k_i \bm{v}_i, \, i = 1,2$. In general, these two eigenframes are different and are rotated by an angle $\theta$, Fig.~\ref{frames}(a), which can be assumed to lie in the interval $\theta \in(-\pi/2, \pi/2]$ since eigenvectors can be flipped. We can express one frame in terms of the other as $\bm{\lambda} = \cos\theta \bm{v}_1 - \sin\theta \bm{v}_2$ and $\bm{\mu} = \sin\theta \bm{v}_1 + \cos\theta \bm{v}_2$. From the definitions of angles $\gamma$ and $\theta$, the angle between the principal curvature direction $\bm{v}_1$ and a microscopic particle direction $\bm{\ell}$ is $\gamma-\theta$, and hence 
\begin{equation}\label{kell}
k_{{\bm{\ell}}} = k_{1} \cos^2(\gamma-\theta) + k_2 \sin^2(\gamma-\theta).
\end{equation}

Denoting the bending rigidity (with units of energy) of a protein by $\kappa_p$, its preferred curvature along the long axis by $\bar{C}$ and its area by $a_p$, we can write its elastic bending energy as  
\begin{equation}\label{Ub}
U^b(\bm{k},\gamma) =\frac{ \kappa_p a_p}{2}  (k_{\bm{\ell}} - \bar{C})^2.
\end{equation}
We note that this energy depends on position (through the principal curvatures $k_1$ and $k_2$) and on the relative orientation between the particle direction and the principal curvature direction. Combining the bending energy with the interaction energy discussed in Section 1 for $N$ proteins, Eq.~\eqref{Int_Potential}, we obtain the free energy
\begin{equation}
\mathcal{F} = - \frac{1}{\beta} \ln {\frac{1}{N!} \int_{\Omega^N} e^{-\beta \left( \Sigma_{1 \leq i < j \leq N} U_{i,j} + \Sigma_{1\leq i \leq N} U^b_i \right)}  \, d \bm{q}_1 \dots d \bm{q}_N}.
\end{equation}
Following a similar mean field approximation and a passage to the continuum limit as in Section 1 and noting that the bending energy of a protein molecule does not depend on the state of other proteins, we arrive at the following expression for the free energy of the system (see Appendix \ref{app})
\begin{equation}\label{curv_tot_ener}
\begin{aligned}
\mathcal{F}[\phi, f] =  & \frac{1}{\beta}\int_{\Gamma} \phi \ln \phi \, d\bm{r} 
+ \frac{1}{\beta} \int_{\Gamma} \phi \left\{ \int_{\mathbb{S}} f \left[ \ln f - \ln g \right]\, d\gamma \right\} \, d\bm{r}     \\ 
 + & \int_{\Gamma} \phi   \int_{\mathbb{S}} f U^b \, d\gamma \, d\bm{r}.
\end{aligned}
\end{equation}
As before, we find the optimal particle angle distribution $f$ by minimizing the free-energy, resulting in an effective energy that will depend on $S$ as before, but now also on $\theta$ and hence on the full nematic tensor $\bm{Q}$. Analogously to before, we introduce the Lagangian
\begin{align}
\mathcal{L} = \int_{\mathbb{S}} f \left[\ln f-\ln g + U^b  \right]d\gamma +  \mu \left(\int_{\mathbb{S}} f \, d\gamma - 1 \right).
\end{align}
Minimization with respect to $f$ requires that 
\begin{align}
0 = \delta_f \mathcal{L} =  \int_{\mathbb{S}} \left[\ln f - \ln g -  \bm{\psi}:\bm{\sigma}  +  U^b + \mu \right] \delta f d\gamma,
\end{align}
where the auxiliary symmetric and traceless tensor $\bm{\psi}$ was defined in Eq.~(\ref{Aux_def}), and hence, with the same argument leading to Eq.~(\ref{PDF_1}), we find that 
\begin{align}\label{PDF_2}
f(\gamma) = & \dfrac{   g(\gamma) \, e^{\bm{\sigma}(\gamma) :  \bm{\psi}} \, e^{-U^b}   }{ \int_{\mathbb{S}^+}   g(\gamma') \, e^{\bm{\sigma}(\gamma') :  \bm{\psi}}  \, e^{-U^b} \,  d\gamma'}
\end{align}
if $\gamma\in \mathbb{S}^+$ and 0 otherwise. 

As before, we express $\bm{\psi}$ in the eigenframe of $\bm{Q}$ as 
\begin{align}
\bm{\psi} = \frac{\psi}{2} (\bm{\lambda}\otimes\bm{\lambda} - \bm{\mu}\otimes\bm{\mu}) + \frac{\bar{\psi}}{2} (\bm{\lambda}\otimes\bm{\mu} + \bm{\mu}\otimes\bm{\lambda}).
\end{align}
However, we cannot make the same argument as before to conclude that $\bar{\psi}= 0$ because, unless $\theta= 0$ or $\theta= \pi/2$, $U^b$ is not an even function of $\gamma$, see Eqs.~(\ref{kell},\ref{Ub}). Recalling Eq.~(\ref{sigma_psi}), we can write the angular probability distribution 
\begin{align}\label{PDF_3}
f(\gamma) = & \frac{ \left[1-\phi\left(c - d S P_2 \right) \right] \, e^{\psi P_2} \,e^{\bar{\psi} \sin\gamma\cos\gamma} \,  e^{-U^b}}{ \int_{\mathbb{S}^+} \left[1-\phi\left(c - d S P_2 \right) \right] \, e^{\psi P_2}\,e^{\bar{\psi} \sin\gamma'\cos\gamma'} \,  e^{-U^b} \,  d\gamma'},
\end{align}
where $P_2$ stands for $P_2(\cos\gamma)$. In Section 1, we used Eq.~(\ref{Order_Def}) to determine $\psi$. Now, however, we need two equations since we also need to determine $\bar{\psi}$. For this, we recall that  $0 = \bm{\lambda}\cdot \bm{Q} \cdot \bm{\mu}$ leading to Eq.~(\ref{condx}). Thus, we have two conditions  
\begin{align}\label{root2}
0 = & 2{\int_{\mathbb{S}^+} P_2 \left[1-\phi\left(c - d S P_2 \right) \right] \, e^{\psi P_2} \,e^{\bar{\psi} \sin\gamma\cos\gamma} \,  e^{-U^b}\, d\gamma} \nonumber \\ & - S  { \int_{\mathbb{S}^+} \left[1-\phi\left(c - d S P_2 \right) \right] \, e^{\psi P_2}\,e^{\bar{\psi} \sin\gamma\cos\gamma} \,  e^{-U^b} \,  d\gamma}, \\
0 = & \int_{\mathbb{S}^+} \sin\gamma \cos\gamma \left[1-\phi\left(c - d S P_2 \right) \right] \, e^{\psi P_2}\,e^{\bar{\psi} \sin\gamma\cos\gamma} \,  e^{-U^b} ~d\gamma, \label{root3}
\end{align}
the second of which was trivially satisfied previously as the integrand is an odd function of $\gamma$ for $\bar{\psi}= 0$ and $U^b=0$. Now, however, these two relations provide a system of nonlinear equations to solve for $\psi$ and $\bar{\psi}$.

Examining the above equations, it is clear that $f(\gamma)$, $\psi$ and $\bar{\psi}$ depend on $\phi$ and $S$, but also on $\theta$, $k_1$ and $k_2$ through $U^b$. Plugging Eq.~(\ref{PDF_3}) into Eq.~(\ref{curv_tot_ener}), we obtain a computable expression of the free energy accounting for the curvature energy of the proteins
\begin{align}\label{Free_Energy_5}
\mathcal{F}&[\phi  ,S, \theta, k_1, k_2] =   \frac{1}{\beta}\int_{\Gamma} \phi \bigg\{  \ln \phi + \frac{S \psi}{2}  \\ & - \ln\int_{\mathbb{S}^+} \left\{1-\phi\left[c - d S P_2(\cos\gamma)   \right] \right\} e^{\psi P_2(\cos\gamma)}\,e^{\bar{\psi} \sin\gamma\cos\gamma} \,  e^{-U^b} d\gamma \bigg\} d\bm{r}.  \nonumber
\end{align}

It is interesting to note that, as mentioned earlier, in the special case that the nematic direction $\bm{\lambda}$ is aligned with one of the principal directions, $\theta = 0$ or $\theta = \pi/2$, then $\bar{\psi} = 0$, $U^b$ becomes an even function of $\gamma$, and hence $f(\gamma)$ is symmetric with respect to the nematic direction. For a general nematic orientation relative to the principal curvatures, however, $f$ is not symmetric about the nematic direction.

\subsection{Optimizing $\theta$}

The free energy in Eq.~(\ref{Free_Energy_5}) can then be minimized with respect to $\theta$ to yield an effective energy 
\begin{align}\label{Free_Energy_6}
\hat{\mathcal{F}}[\phi  ,S, k_1, k_2] = \min_{\theta\in(-\pi/2,\pi/2)} \mathcal{F}[\phi  ,S, \theta, k_1, k_2].
\end{align}
This process identifies the energetically optimal nematic orientation relative to the curvature of the surface. To do that, we make $\mathcal{L}$ stationary with respect to $\theta$ to find 
\begin{align}
0 &= \int_{\mathbb{S}^+} g(\gamma) e^{\bm{\sigma}:\bm{\psi}} e^{-U^b}\frac{\partial U^b}{\partial \theta}\,d\gamma.
\end{align}
Expanding the last term in the integral, we find 
\begin{align}\label{root4}
0 = \int_{\mathbb{S}^+} & g(\gamma) e^{\bm{\sigma}:\bm{\psi}} e^{-U^b} 
\left[k_1 \cos^2(\gamma-\theta) + k_2 \sin^2(\gamma-\theta) - \bar{C} \right] \\ & (k_1 - k_2) \cos(\gamma-\theta)\sin(\gamma-\theta)
  \,d\gamma.\nonumber
\end{align}
This equation, together with Eqs.~(\ref{root2},\ref{root3}), provides a system of three nonlinear equations for three unknowns, $\psi$, $\bar{\psi}$ and $\theta$. For a sphere, $k_1 = k_2$,  this equation is an identity showing that any direction is equally possible. Suppose that  $\bar{\psi} = 0$ and $\theta = 0$. In this case, $U^b$ is an even function of $\gamma$,  Eqs.~(\ref{root3},\ref{root4}) are identically satisfied, and Eq.~(\ref{root2}) provides an equation for $\psi$. Thus, there is always a solution with  $\bar{\psi} = 0$ and $\theta = 0$ but in general there may be others and their relative stability must be examined to select the ground state.

\begin{figure*}[hbtp!]
\centering
\includegraphics[width=7.5in]{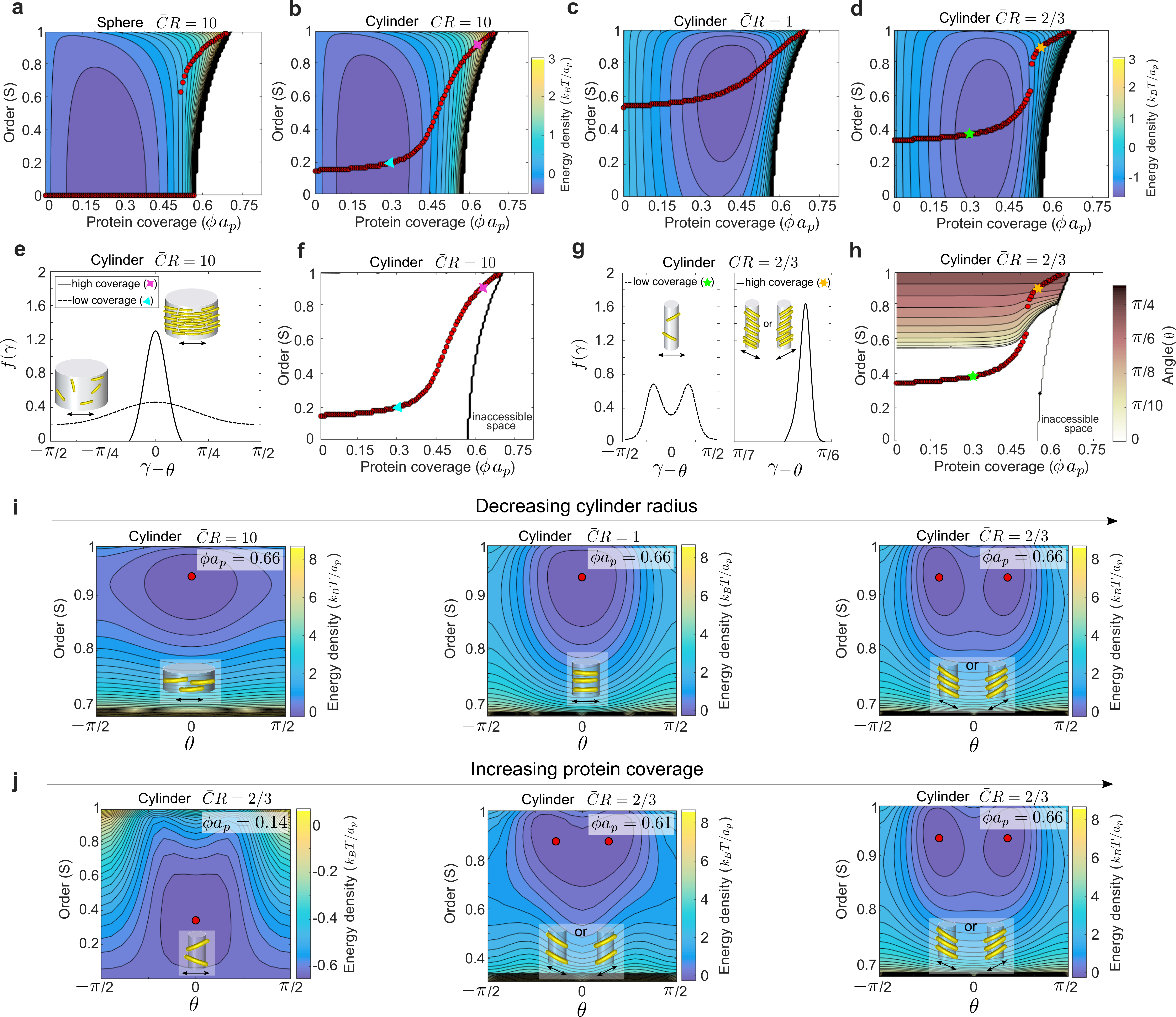}
\caption{\label{Ener_land_curv} (a)-(d) Energy density contours as a function of density and order on spherical and cylindrical surfaces of different radii, $R=150, 150, 15$  and 10 nm, with $\bar{C}= 15$ nm. Red dots denote stable states, which minimize the free-energy for a given protein coverage. (e,g) Angular probability distribution  $f(\gamma)$ plotted against $\gamma-\theta$, i.e. against the angle of a particle relative to the direction of maximum curvature of the cylinder $\bm{v}_1$, Fig.~\ref{frames}. The inset pictorially illustrates the state of the system, where the double-ended arrow indicates the nematic direction. (f,h) Protein net orientation expressed as the angle $\theta$ between the nematic direction and $\bm{v}_1$ as a function of density and order for the cylindrical surfaces in (b) and (d). In (f), $\theta=0$ everywhere. (i) Energy landscapes in the ($\theta-S$) plane for high protein coverage ($\phi a_p = 0.66$) and cylinders of decreasing radius. (j) Energy landscape in the ($\theta-S$) plane for a thin cylinder ($\bar{C}R=2/3$) and varying coverage.}
\end{figure*}

\subsection{Free-energy landscapes on curved surfaces}

Energy density landscapes exhibiting the isotropic-to-nematic transition for proteins on spherical and cylindrical surfaces are shown in Fig.~\ref{Ener_land_curv} (a-d) using the expression given by Eq.~\eqref{Free_Energy_5} and minimizing the energy with respecto to $\theta$ (the angle between the net orientation of proteins and the maximum curvature direction) as described in the previous section. We depict stable equilibrium states minimizing the free energy for a given protein coverage with respect to $S$ and $\theta$  with red dots. For a sphere, Fig.~\ref{Ener_land_curv}(a), the isotropic curvature does not bias alignment along any specific direction and hence the phase transition is solely driven by entropic interactions. In fact, examining Eq.~(\ref{curv_tot_ener}), it is clear that since $U^b$ does not depend on orientation, the last bending term in the free-energy density is simply linear in $\phi$ and hence does not alter the path of minimizers marked by red dots. As a result, the system shows the same discontinuous transition upon crowding as in the planar case. 

On anisotropically curved surfaces such as cylinders with radius $R$, proteins are biased to orient along specific directions to favorably adapt their curvature to that of the underlying surface,  Fig.~\ref{Ener_land_curv}(b-d). This creates a competition between a curvature-dependent bias and the entropic part of the free-energy, which leads to partial order (finite $S$) even in the dilute limit. Furthermore, this curvature bias changes the character of the isotropic-to-nematic transition, which now becomes continuous. Our model not only provides the free-energy landscape as a function of $\phi$ and $S$ but also the nematic orientation relative to the direction of maximum curvature of the cylinder $\bm{v_1}$ (Fig.~\ref{frames}) given by $\theta$ and represented in Fig.~\ref{Ener_land_curv}(f,h), and the distribution of protein orientations $f(\gamma)$,  which we represent relative to $\bm{v_1}$, i.e.~against $\gamma-\theta$, Fig.~\ref{Ener_land_curv}(e,g). Figure \ref{Ener_land_curv}(f) illustrates the observation that for $\bar{C}R \ge 1$ the optimal nematic orientation is always that of maximum curvature of the cylinder, $\theta = 0$. Figure \ref{Ener_land_curv}(e) shows the angular distribution for two values of protein coverage marked in (b). Both distributions are unimodal and symmetric about the direction given by $\bm{v_1}$ but as coverage increases, order increases as well and the distribution becomes more localized and compactly supported. 

For cylinders with higher curvature than that of proteins, $\bar{C}R < 1$, the situation is more complex since now proteins aligned with $\bm{v_1}$ will be bent beyond their spontaneous curvature whereas proteins forming an angle $\alpha$ with $\bm{v_1}$ given by $\cos^2\alpha = \bar{C}R$, see Eq. ~\ref{kell}, will store no elastic energy. The isotropic-to-nematic transition becomes discontinuous again, Fig.~\ref{Ener_land_curv}(d), and the nematic direction is along $\bm{v_1}$ for low $S$ but $\theta$ becomes different from zero for larger order, Fig.~\ref{Ener_land_curv}(h). Interestingly, at low densities when $\theta = 0$, the angular distribution is bimodal, broad, and symmetric about $\bm{v_1}$, indicating a state where proteins are disordered but preferentially adopt orientations forming a finite angle with the direction of maximum curvature, which is too curved compared to the protein curvature. This detailed information is lost if the nematic state is described in terms of a moment of $f$ such as the nematic tensor $\bm{Q}$ or equivalently $S$ and $\theta$ alone, rather than in terms of the full distribution. For high density, we find that the system adopts a non-symmetric, very narrow and compactly supported distribution (or a symmetry-related distribution), indicative of a nematic state with nematic direction forming a finite angle with $\bm{v_1}$, consistent with the fact that F-BAR proteins at high coverages adopt increasingly helical arrangements on increasingly thinner tubes \cite{Frost2008}.

The symmetry-breaking transition for thin tubes at high coverages can be nicely examined through free-energy maps at given coverage and radius as a function of $\theta$ (the angle between nematic direction and $\bm{v}_1$) and $S$. On the one hand, we can observe how at fixed high-coverage and as the cylinder radius decreases, a single minimum given by $\theta=0$ splits into two minima given by $\theta=\pm\theta_0$ when $\bar{C}R < 1$, Fig.~\ref{Ener_land_curv}(i). Similarly, given a high-curvature cylinder, as coverage increases the optimal nematic direction switches from being aligned with $\bm{v}_1$ to adopting either one of two symmetry-related orientations, Fig.~\ref{Ener_land_curv}(j).

\begin{figure*}[hbtp!]
\centering
\includegraphics[width=7.5in]{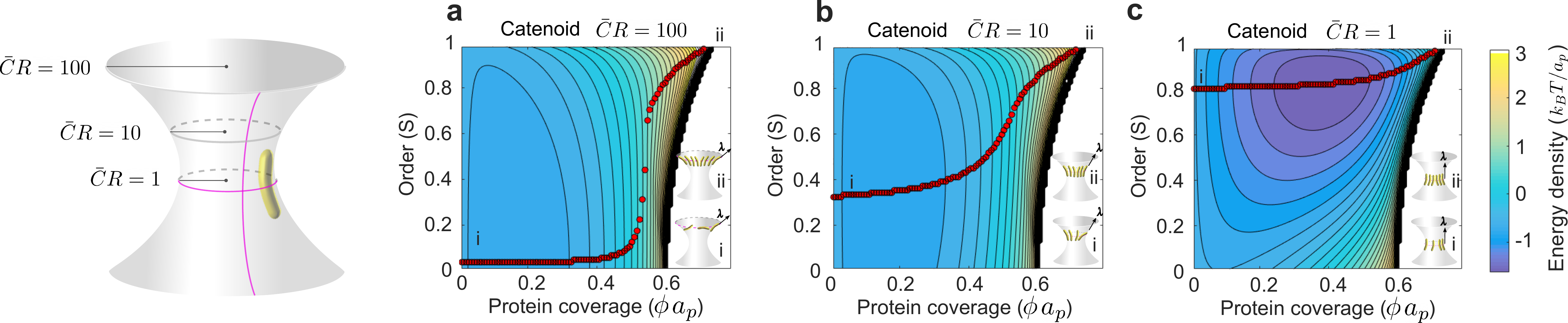}
\caption{\label{Ener_land_catenoid} (a)-(c) Energy density contours as a function of density and order 
 on negatively curved surfaces of zero mean curvature, where the principal curvatures are $k_{1,2} = \pm 1/R$, for different values of $R$. By way of illustration, we identify these curvatures with different positions on a catenoid surface.}
\end{figure*}

We finally note that the model proposed here can be used to quantify the free-energy of curved elongated particles on other surfaces, such as those of negative curvature. Figure \ref{Ener_land_catenoid} shows the isotropic-to-nematic transition on different regions of a catenoid, characterized by having zero mean curvature and negative Gaussian curvature. The results are similar to those on cylinders, albeit with a larger bias towards nematic states, compare Fig.~\ref{Ener_land_curv}(b,c) and Fig.~\ref{Ener_land_catenoid}(b,c).

\begin{figure*}[hbtp!]
\centering
\includegraphics[width=6.5in]{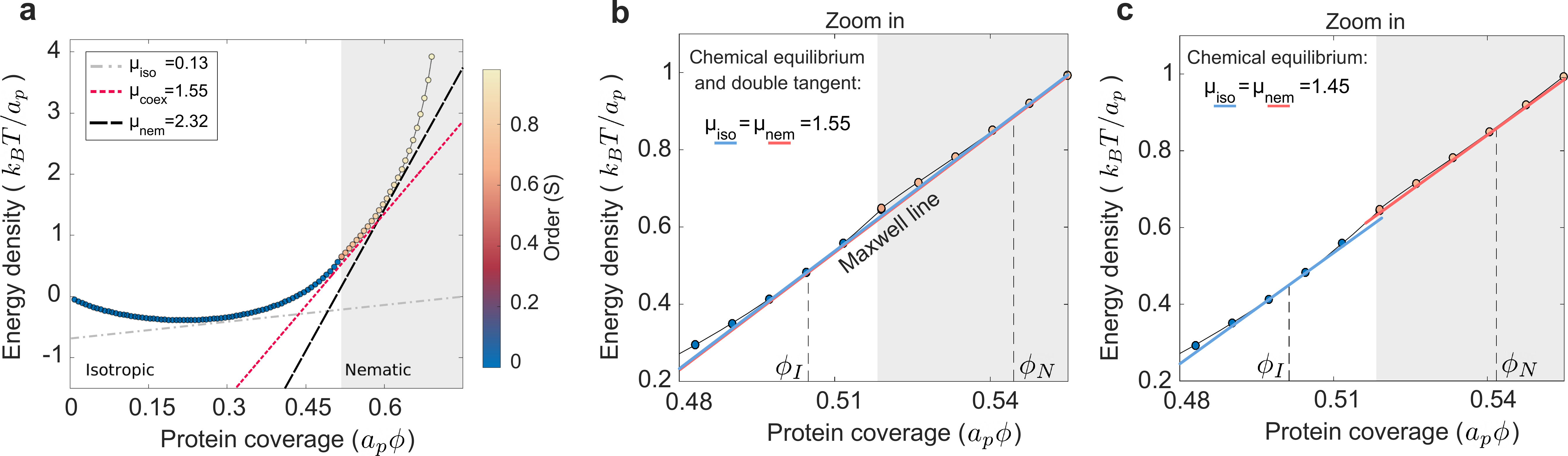}
\caption{\label{Coexistence_Ellipses} (a) Lowest energy density as a function of protein coverage for ellipses on a flat membrane, i.e.~the energy along the minimum-energy path marked with red dots in Fig.~\ref{Energy_ellipses}. The color code represents order, highlighting the parts of the energy landscape corresponding to isotropic and to nematic phases. The slopes of the tangent lines represent the rate of change of energy density with respect to protein coverage, i.e.~the chemical potential. The red line is doubly tangent (a Maxwell line) to the isotropic and nematic branches and represents a situation of coexistence in which protein number is fixed, see zoom in (b). If proteins can be exchanged with a bulk solution where they have a given chemical potential, then coexistence of isotropic and nematic phases does not require the double tangency constraint (c).}
\end{figure*}

\section{Coexistence of isotropic and nematic phases}

Coarse-grained simulations suggest the possibility of coexistence between isotropic and nematic phases \cite{Noguchi2016}. To examine such coexistence using our theory, we first consider the situation of a flat membrane. Figure \ref{Coexistence_Ellipses}(a) shows the landscape of minimum free energy as a function of protein coverage, which as discussed earlier and shown here with the color representing order, has an isotropic and a nematic branch. The slope of the energy density as a function of protein coverage is precisely the chemical potential of proteins in a given state. For coexistence of isotropic and nematic phases in equilibrium, the chemical potentials of the two phases should be equal. If the number of proteins populating these two phases is fixed with average density $\bar{\phi}$, then coexistence additionally requires the double tangency condition, see Fig.~\ref{Coexistence_Ellipses}(a,b) and the line with slope $\mu_{\text{coex}}$ following the Maxwell construction. Thus, under these conditions, coexistence is possible only when $\phi_I < \bar{\phi} < \phi_N$. When the membrane can exchange proteins with a bulk solution with chemical potential $\mu_b$, the double tangency condition is no longer required and coexistence requires simply that $\mu_b = \mu_I = \mu_N$, see Fig.~\ref{Coexistence_Ellipses}(c). This slightly relaxes the possibility of coexistence but the figure shows that it can only occur in a rather narrow range of densities. 

\begin{figure}[hbtp!]
 \centering
\includegraphics[width=3in]{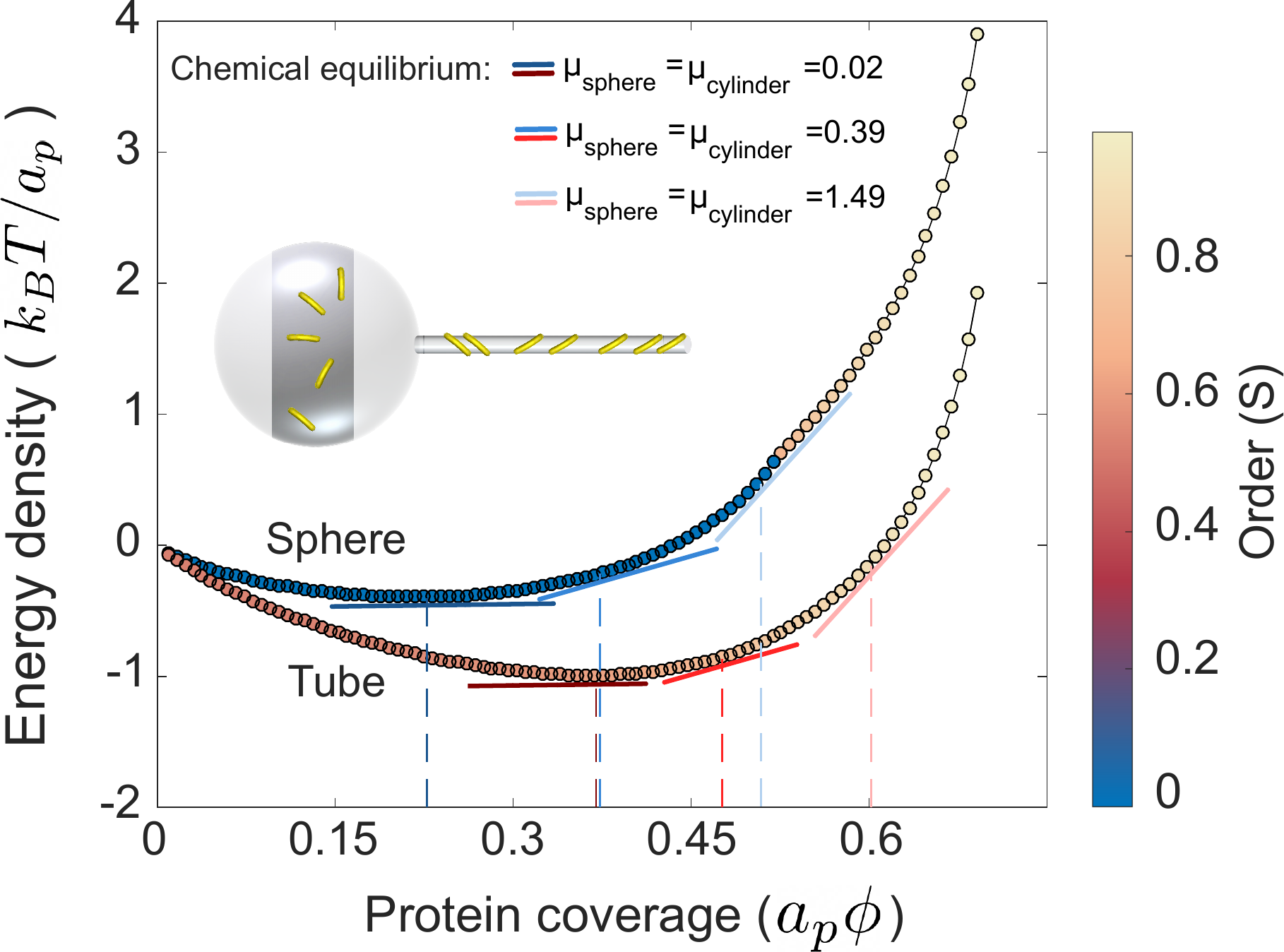}
\caption{\label{Coexistence_Sphere_Tubes} Energy landscape for a sphere with radius $R_s = 10/\bar{C}$ and a tube with radius $R_t =2/\bar{C}$. Three different isotropic-nematic states of coexistence in an ensemble in which proteins can be exchanged with a bulk solution are highlighted by pairs of tangents with the same slope.}
\end{figure}

Since as discussed earlier the energy landscape on spheres is that of a planar surface with a tilt proportional to $\phi$, the conditions for coexistence are similar. On tubes, however, there exists essentially no isotropic phase, and as illustrated in Fig.~\ref{Coexistence_Sphere_Tubes} the energy is convex in $\phi$, leaving no room for coexistence. Yet, as suggested by experiments where thin tubes are pulled off giant vesicles and exposed to a solution with BAR proteins \cite{Sorre2012}, it is reasonable to expect isotropic-nematic coexistence in cylinder-sphere systems in equilibrium. Indeed, for moderate coverages, spheres adopt an isotropic state whereas thin-enough tubes are in a significantly nematic state. It is thus possible to find infinitely many equilibrium states of coexistence over a broad range of bulk chemical potentials, Fig.~\ref{Coexistence_Sphere_Tubes}, in all of which area coverage and order are higher on the tube.

\section{Conclusions}

Curved proteins on membranes are responsible for many biological functions, which rely on the mechanisms of curvature sensing and generation. When these proteins are elongated, such as those containing BAR domains, their physics crucially depend on their orientation. Here, extending the work of \cite{Nascimento_2017}, we have developed a mean-field density functional theory connecting physics from a micro- to a mesoscale to evaluate the free-energy of elongated and curved proteins on a curved membrane applicable to large protein coverage. The free-energy landscape is expressed in terms of the net orientation of proteins relative to the principal curvature directions ($\theta$), of the classical order parameter $S$, and of the number density $\phi$, and it depends on the aspect ratio and intrinsic curvature of these proteins, on their bending rigidity, and on the second fundamental form of the membrane. In addition to the free-energy landscape, the theory provides the orientational probability distribution of proteins. 

We have shown that, while on planar surfaces and spheres the system exhibits a density-dependent discontinuous isotropic-to-nematic transition, this transition is continuous on surfaces with anisotropic curvature such as cylinders or catenoids. We have shown that anisotropic curvature biases the system towards a slightly nematic state even at low protein concentrations. When the curvature of cylindrical membranes is higher than that of the proteins, then the orientational distribution becomes bimodal at low densities and asymmetric with respect to the principal direction of curvatures at high densities.  Our theory has also allowed us to examine the coexistence of isotropic and nematic phases under different conditions.

Our theory provides physical rules to understand the state curved and elongated proteins on surfaces of given curvature. However, it does not say anything about how the proteins, with a given density or orientational distributions, affect the shape of the underlying membrane. This situation is examined experimentally and computationally elsewhere \cite{LeRoux20}, where the present model is coupled with one of membrane dynamics.


\section*{Conflicts of interest}
There are no conflicts to declare.

\section*{Acknowledgements}
This work was supported by the Spanish Ministry of Economy and Competitiveness/FEDER (BES-2016-078220 to C.T.), the European Commission (H2020-FETPROACT-01-2016-669 731957), the European Research Council (CoG-681434 to M.A.), the Generalitat de Catalunya (2017-SGR-1602 to P.R-C., 2017-SGR-1278 to M.A.), the prize ``ICREA Academia'' for excellence in research to P.R-C. and to M.A., and Obra Social ``La Caixa''. IBEC and CIMNE are recipients of a Severo Ochoa Award of Excellence from the MINECO.

\appendix
\section{Mean field free-energy of bendable proteins on a curved membrane}
\label{app}

The free-energy of $N$ proteins interacting with the pairwise potential as mentioned in Eq.~\eqref{Int_Potential} and also with the underlying curved surface as given by Eq.~\eqref{Ub} is
\begin{equation}
\mathcal{F}^* = - \frac{1}{\beta} \ln {\frac{1}{N!} \int_{\Omega^N} e^{-\beta \left( \Sigma_{1 \leq i < j \leq N} U_{i,j} + \Sigma_{1\leq i \leq N} U^b_i \right)}  \, d \bm{q}_1 \dots d \bm{q}_N}.
\end{equation}
As shown in \cite{Gartland2010, Nascimento_2017}, the above free energy can be approximated by the mean field energy 
\begin{align}
\mathcal{F}^* \approx \mathcal{F}  & =  - \frac{1}{\beta} \ln {\frac{1}{N!}  \left< \int_{\Omega} e^{-\beta \sum_{2 \leq j \leq N} U_{1,j}}  e^{-\beta U^b_1}\, d \bm{q}_1 \right>}^N \\
& = - \frac{1}{\beta} \ln {\frac{1}{N!}  \left( \int_{\Omega} \left< e^{-\beta \sum_{2 \leq j \leq N} U_{1,j}}  e^{-\beta U^b_1}\right> \, d \bm{q}_1 \right)}^N,
\end{align}
where in the second line we have changed the order of integration and where the ensemble average 
\begin{equation}
\left< f \right> = \int_{\Omega^{N-1}} f \: p(\bm{q}_2, \dots,  \bm{q}_{N}) \, d \bm{q}_2 \dots d \bm{q}_N,
\end{equation}
is with respect to the probability distribution of $N-1$ particles given by
\begin{equation}
p(\bm{q}_2, \dots , \bm{q}_{N}) = \frac{ e^{-\beta \left( \Sigma_{2 \leq i < j \leq N} U_{i,j} + \Sigma_{2\leq i \leq N} U^b_i \right)}}{\int_{\Omega^{N-1}}  e^{-\beta \left( \Sigma_{2 \leq i < j \leq N} U_{i,j} + \Sigma_{2\leq i \leq N} U^b_i \right)}   \, d \bm{q}_2 \dots d \bm{q}_{N}}.
\end{equation}
The mean free approximation is an  upper bound to the exact free energy.

Unlike the hard-core repulsion energy, which  depends on the configuration of other proteins, the bending energy of a protein molecule does not depend on the configuration of other proteins. We can thus  write
\begin{equation}
\begin{aligned}
\mathcal{F} & =  - \frac{1}{\beta} \ln {\frac{1}{N!}  \left( \int_{\Omega} \left< e^{-\beta \sum_{2 \leq j \leq N} U_{1,j}} \right>  e^{-\beta U^b_1} \, d \bm{q}_1 \right)}^N \\
& =  - \frac{1}{\beta} \ln {\frac{1}{N!} \left( \int_{\Omega} [ 1 - W(\bm{q}_1)]  e^{-\beta U^b_1} \, d \bm{q}_1 \right)^N}
\end{aligned}
\end{equation}
or equivalently
\begin{equation}
\mathcal{F} = -\frac{1}{\beta} \ln \frac{1}{N!} \left( \int_{\Omega} [ 1 - W(\bm{q})]  e^{-\beta U^b (\bm{q})} \, d \bm{q} \right)^N
\end{equation}
with $ W(\bm{q})$ as defined before in Eq.~\eqref{W_excluded}. Following \cite{Nascimento_2017}, we discretize the phase space in subdomains  $\Omega = \cup_i \Omega_i$, each with  $N_i$ particles, and obtain free energy within this domain $\mathcal{F}_i$ after using Stirling's approximation as
\begin{equation}
\mathcal{F}_i = - \frac{1}{\beta} \ln \left(  {\frac{1}{N_i}  \int_{\Omega_i} }  [ 1 - W(\bm{q}_i)]  e^{-\beta U^b (\bm{q}_i)} \, d \bm{q}_i \right)^{N_i}.
\end{equation}
Thus, the total free energy is given by
\begin{equation}
\mathcal{F} = \sum_{i} \mathcal{F}_i = - \frac{1}{\beta} \ln \prod_{i}  \left(  {\frac{1}{N_i}  \int_{\Omega_i} }  [ 1 - W(\bm{q}_i)]  e^{-\beta U^b (\bm{q}_i)} \, d \bm{q}_i \right)^{N_i}.
\end{equation}
Assuming $N_i \approx {\rho} (\bm{q}_i) \Delta \bm{q}_i$ and passing onto the continuum limit, we obtain 
\begin{equation}
\begin{aligned}
\mathcal{F} & =  \frac{1}{\beta} \int_{\Omega} {\rho}(\bm{q}) \ln {{\rho}(\bm{q})} \, d\bm{q} -  \frac{1}{\beta} \int_{\Omega} \rho(\bm{q}) \ln [1 - W(\bm{q})] \, d\bm{q}\\
& + \int_{\Omega} \rho(\bm{q}) U^b (\bm{q}) \, d\bm{q}.
\end{aligned}
\end{equation}
Further, separating the particle density $\rho$ into spatial and orientational components as mentioned in Eq.~\eqref{separation}, we obtain the expression for free energy in Eq.~\eqref{curv_tot_ener}



\balance


\bibliography{Bib_Orientational_Order} 
\bibliographystyle{rsc} 

\end{document}